\documentclass[]{jfm}
\usepackage{graphicx,scalerel}
\newcommand\sbullet[1][.5]{\mathbin{\ThisStyle{\vcenter{\hbox{%
 \scalebox{#1}{$\SavedStyle\bullet$}}}}}%
}

\usepackage{newtxtext}
\usepackage{newtxmath}
\usepackage{natbib}
\usepackage{enumitem}
\usepackage{hyperref}
\usepackage{tikz}
\usepackage{amssymb}
\usepackage{mathtools}
\newcommand\cf{c.f.\ }
\newcommand\ie{i.e.\ }
\hypersetup{
 colorlinks = true,
 urlcolor = blue,
 citecolor = blue,
 linkcolor = blue
}

\newcommand{\RomanNumeralCaps}[1]
\linenumbers

\newcommand\Tay{\mbox{\textit{Ta}}} 
\newcommand\Stk{\mbox{\textit{St}}} 
\newcommand\Nus{\mbox{\textit{Nu}}} 

\title{Strong alignment of prolate ellipsoids in Taylor--Couette flow }

\author[M.P.A. Assen \& others]%
{Martin P.A. Assen$^1$ \thanks{Email address for correspondence: m.p.a.assen@utwente.nl, c.s.ng@utwente.nl,\newline d.lohse@utwente.nl, verzicco@uniroma2.it}, Chong Shen Ng$^1$, Jelle B. Will$^1$, Richard J. A. M. Stevens$^1$, Detlef Lohse$^{1,4}$ and Roberto Verzicco$^{1,2,3}$}

\affiliation{$^1$Physics of Fluids Group and Max Planck Center Twente, MESA+ Institute and J. M. Burgers Centre
for Fluid Dynamics, University of Twente, P.O. Box 217, 7500AE Enschede, The Netherlands
\\[\affilskip]
$^2$Dipartimento di Ingegneria Industriale, University of Rome ‘Tor Vergata’,\\ Via del Politecnico 1, Roma 00133, Italy
\\[\affilskip]
$^3$Gran Sasso Science Institute - Viale F. Crispi, 7 67100 L'Aquila, Italy
\\[\affilskip]
$^4$Max Planck Institute for Dynamics and Self-Organization, Am Fassberg 17, 37077 G\"ottingen, Germany
}

\begin{document}
\maketitle
\begin{abstract}
We report on the mobility and orientation of finite-size, neutrally buoyant prolate ellipsoids (of aspect ratio $\Lambda=4$) in Taylor--Couette flow, using interface resolved numerical simulations. The setup consists of a  particle-laden flow in between a rotating inner and a stationary outer cylinder. The flow regimes explored are the well known Taylor vortex, wavy vortex, and turbulent Taylor vortex flow regimes. We simulate two particle sizes  $\ell/d=0.1$ and $\ell/d=0.2$,  $\ell$ denoting the particle major axis and $d$ the gap-width between the cylinders. The volume fractions are $0.01\%$ and $0.07\%$,  respectively. The particles, which are initially randomly positioned, ultimately display characteristic spatial distributions which can be categorised into four modes. Modes $(i)$ to $(iii)$ are observed in the Taylor vortex flow regime, while mode ($iv$) encompasses both the  wavy vortex, and turbulent Taylor vortex flow regimes.  Mode $(i)$ corresponds to stable orbits away from the vortex cores. 
Remarkably, in a narrow $\Tay$ range, particles get trapped in the Taylor vortex cores (mode ($ii$)). Mode $(iii)$ is the transition when both modes $(i)$ and $(ii)$ are observed. For mode $(iv)$, particles distribute throughout the domain due to flow instabilities. All four modes show characteristic orientational statistics. The focus of the present study is on mode $(ii)$.
We find the particle clustering for this mode to be size-dependent, with two main observations. Firstly, particle agglomeration at the core is much higher for $\ell/d=0.2$ compared to $\ell/d=0.1$. Secondly, the  $\Tay$ range for which clustering is observed depends on the particle size. 
For this mode $(ii)$ we observe particles to align strongly with the local cylinder tangent. The most pronounced particle alignment is observed for $\ell/d=0.2$ around $\Tay=4.2\times10^5$. 
This observation is found to closely correspond to a minimum of axial vorticity at the Taylor vortex core ($\Tay=6\times10^5$) and we explain why.
\end{abstract}


\section{Introduction}
Particle-laden flows are ubiquitous both in nature and industrial applications. For example, in rivers, where the deposition of large grains can influence the solutal and nutrient exchange processes \citep{Ferdowsi2017}. Or in oceans, where prediction for the accumulation of large plastic debris remains a topic of ongoing research \citep{cozar2014}. In industrial applications the accumulation of particles in turbo-machineries can reduce the efficiency and even damage rotor or stator blades \citep{hamed2006}. Another example is in the paper-making industry, where the orientation of the fibres in the pulp suspension determines the mechanical strength of the final product \citep{lundell2011}. Given the importance of particle-laden flows, understanding phenomena such as transport and clustering is key to optimise engineering applications.

Studies of particle-laden flows \citep{tos09,sal09,bal10,mathai2020bubbly}, in general, have shown a rich phenomenology and can be broadly grouped into two categories. The first  focuses on the flow response due to the presence of the particles. Most frequently, these particles are assumed to be spherical. Examples of these studies include investigations into the influence of particles on turbulent structures \citep{wang2018,ardekani2019,ramesh2019,yu2021}, drag \citep{ardekani2017,wang2021,andersson2012} and the turbulent energy budget \citep{peng2019,Olivieri2020}. The second category focuses on explaining the dynamics of particles in these flows themselves. This becomes particularly interesting for nonspherical particles. For instance, on how shape influences particle rotation \citep{Byron2015,zhao2015} or how particles cluster and preferentially align \citep{henderson2007,ni2014,uhlmann2017,Voth2017,Majji2018,Bakhuis2019}.

A review of the existing literature reveals that for spherical particles the particle dynamics in wall-bounded shear flows is reasonably well-understood. However, apart from the recent work \cite{Bakhuis2019}, little is yet known about the interactions of nonspherical particles in turbulent flows with curvature effects. The objective of this work is to fill this gap. Questions we ask are, for instance, how do nonspherical particles respond to shear flow with large-scale flow structures (due to different lift/drag forces), and do they exhibit preferential clustering or alignment? Another unresolved question regards the relation between the particle size compared to that of the flow features in the fluid phase. In particular, we want to work out the underlying physics.

As a very well controlled shear flow, we choose Taylor--Couette flow (TC) and then study the two-way coupled dynamics of finite-size inertial anisotropic particles in this flow.
 TC is convenient for the following reasons: First, the flow regimes of TC  are well understood and documented \citep{andereck1986,fardin2014,Grossmann2016}. Second, it is a closed system with exact balances \citep{eckhardt2007} that is very accessible, both numerically and experimentally, due to its relatively simple geometry and symmetries. 

To fully resolve the motion of the ellipsoidal particles and their two-way interaction with the surrounding fluid, we employ the Immersed Boundary Method (IBM) using the moving-least squares algorithm \citep{vanella2009,detullio2016,Spandan2017}. IBM is computationally advantageous for this application since the underlying  grid is fixed and no computationally expensive remeshing is needed \citep{Breugem2012}. Secondly, it is straightforward in IBM to vary the size of the particles. The disadvantage of IBM is that inter-particle and particle-wall collisions need to be modelled. Here, we adopt the collision model of \cite{costa2015}, which has been widely tested and employed in studies for particle-laden flows \citep[e.g.][]{ardekani2017, yousefi2020}.

The present work is structured as follows. In \S\,\ref{sec:Setup}, we describe the TC setup and give an overview of the investigated flow regimes. In \S\,\ref{sec:govEqAndNumMethod}, we present the details of the numerical method describing the dynamics of the fluid and particles.
In \S\,\ref{sec:SpatialDistributions}, we present the spatial distributions of particles  and categorise them into modes ($i$) to ($iv$). Then, we compare the modes for the two simulated particle sizes via the joint probability density function (pdf)  of the particle radial position versus the driving of the TC system. In \S\,\ref{sec:AngularStatistics}, we investigate the particle orientations with respect to the local cylinder tangent for the categorised modes. Here, we observe a strong particle alignment, which  we correlate to the axial vorticity of the Taylor vortices.
Finally, in \S\,\ref{sec:conclusion}, we summarise our results. 

\begin{figure}
	\centering
		\begin{tikzpicture}
		\tikzstyle{every node}=[font=\normalsize\selectfont]
		\node[anchor=south west,inner sep=0] (image) at (0,0) {\includegraphics[height=0.33\textwidth]{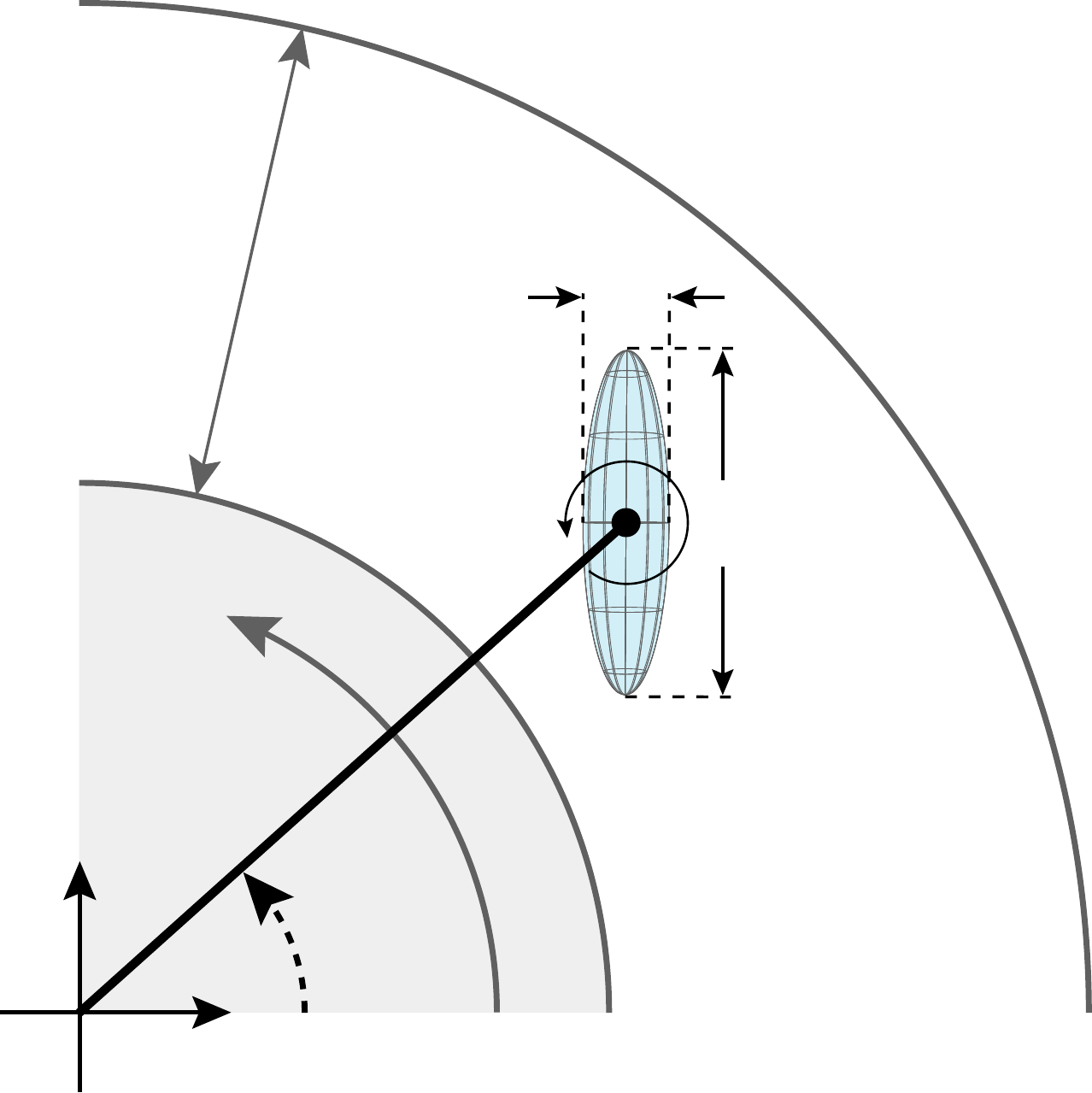}};
		\begin{scope}[x={(image.south east)},y={(image.north west)}]
		\node at (0.66,0.52) {$\ell$};
		\node at (0.46,0.52) {$\boldsymbol{\omega }_p$};
		\node at (0.575,0.725) {$b$};
		\node at (0.02,0.55) {$r_i$};
		\node at (0.02,0.99) {$r_o$};
		\node at (0.17,0.77) {$d$};
		\node at (0.2,0.05) [below] {$x$};
		\node at (0.02,0.23) [below] {$y$};
		\node at (0.15,0.27) [below] {$r_p$};
		\node at (0.34,0.2) [below] {$\varphi_p$};
		\node at (0.16,0.43) {$\omega_i$};
		\node at (0.,1.1) {$(a)$};
		\end{scope}
		\end{tikzpicture}
	\hspace{10pt}
		\begin{tikzpicture}
	\tikzstyle{every node}=[font=\normalsize\selectfont]
	\node[anchor=south west,inner sep=0] (image) at (0,0) {\includegraphics[height=0.33\textwidth]{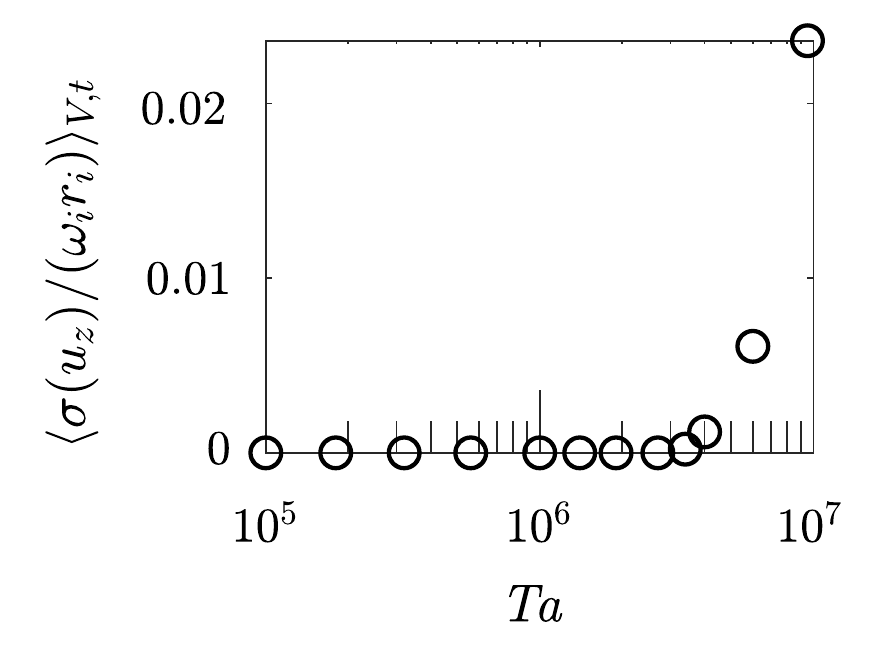}};
	\begin{scope}[x={(image.south east)},y={(image.north west)}]
	\node at (0.,1.1) {$(b)$};
	\end{scope}
	\end{tikzpicture}
	\caption{ $(a)$ Schematic of the TC configuration and geometrical definitions of the particle (not to scale).  $(b)$ The standard deviation of the normalised vertical velocity averaged over the domain and time versus $\Tay$.}
	\label{fig:tc}
\end{figure}
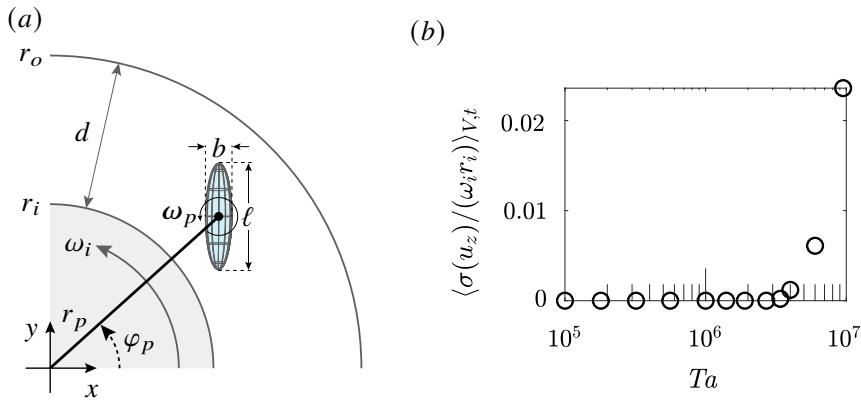

\section{Taylor--Couette setup in the Taylor vortex flow regime} \label{sec:Setup}
The TC setup, as employed here, comprises a confined fluid layer between a coaxially rotating inner and a fixed outer cylinder (see figure \ref{fig:tc}$a$ for a schematic). The dimensionless parameters characterising this system are the ratio of the inner radius $r_i$ and outer radius $r_o$ of the cylinders, \ie\,$\eta \equiv r_i/r_o$, the aspect ratio of the domain $\Gamma \equiv L/d$, and the Reynolds number of the inner cylinder, $\Rey_i \equiv r_i\omega_id/\nu$. Here, $L$ denotes the axial length of the cylinders, $d \equiv r_o - r_i$ the gap width, $\omega_i$ the angular velocity of the inner cylinder, and $\nu$ the kinematic viscosity of the fluid. We set $\Gamma =2\pi/3$ to allow one pair of Taylor vortices to fit within the domain \citep{Ostilla-Monico2015} and $\eta=5/7$ to match the experimental T$^3$C setup \citep{Bakhuis2019}. No-slip and impermeability boundary conditions are imposed on both cylinder walls. In the azimuthal and axial directions, periodic boundary conditions are used. We employ a rotational symmetry of order 6 to reduce computational cost such that the streamwise aspect ratio of our simulations $L_\varphi/d=(2\pi r_i/6)/d=2.62$. The resulting streamwise domain length is sufficient to capture the mean flow statistics \citep{Ostilla-Monico2015}.

\begin{figure}
 			\centering
 			\vspace{5pt}
 	 	\begin{tikzpicture}
 	\tikzstyle{every node}=[font=\normalsize\selectfont]
 	\node[anchor=south west,inner sep=0,rotate=-90] (image) at (0,0) {\includegraphics[height=0.15\textwidth]{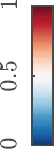}};
 	 	\begin{scope}[x={(image.south east)},y={(image.north west)}]
 	\node [above] at (-0.2,0.5) { $u_{\varphi}$};
 	\end{scope}
 	\end{tikzpicture}
 	\vspace{10pt}\\
 	\includegraphics[width=0.99\linewidth]{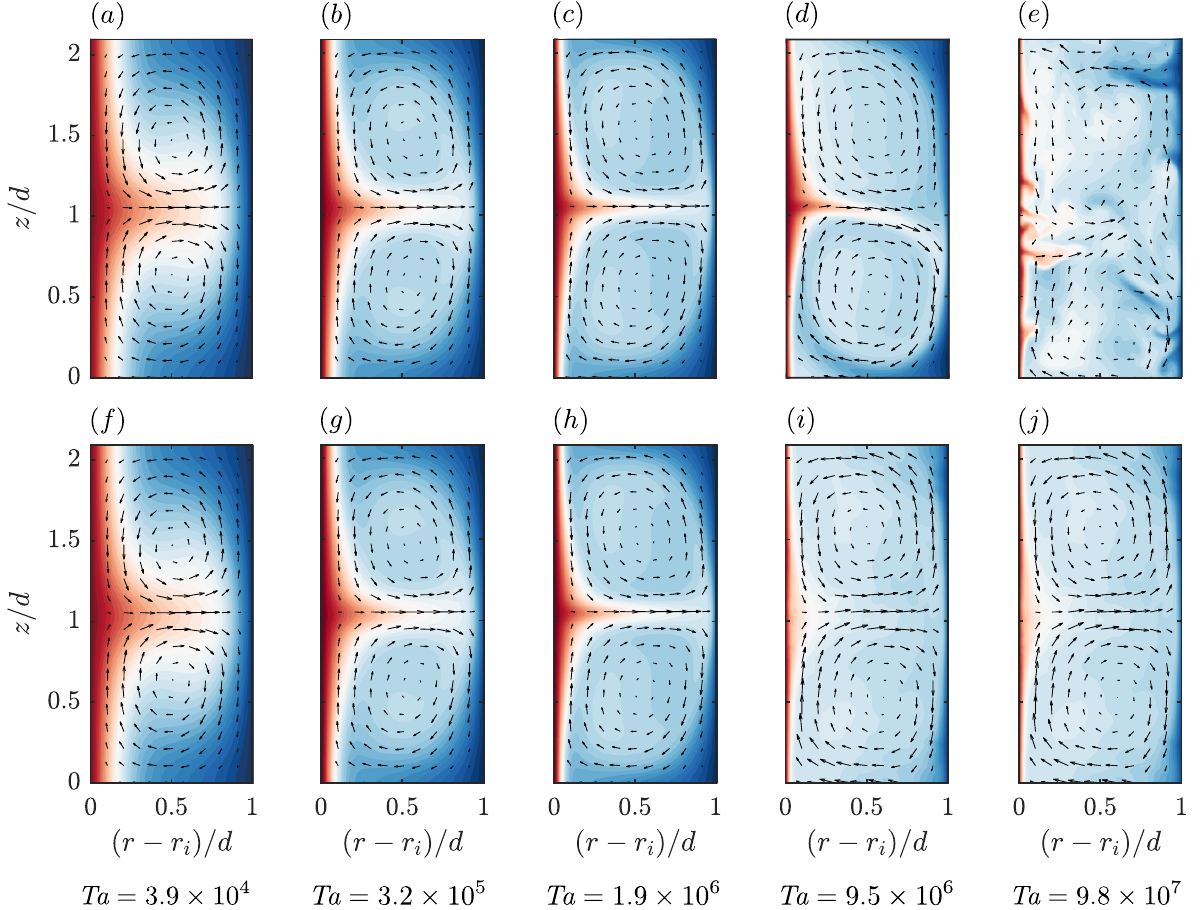}
 	\caption{{\it(a)-(e)} Instantaneous snapshots of the  azimuthal flow field (arrows denote $u_r,u_z$)  for various $\Tay$. {\it (f)-(j)} Corresponding 
 	time-averaged velocity fields. Here, the flow regimes are {\it (a)-(c)}  Taylor vortex flow, {\it (d)} wavy vortex flow,  and {\it(e)} turbulent Taylor vortex flow. } 
 	\label{fig:vel_ta2e6}
 \end{figure}

The control parameter for the TC flow is $\Rey_i$ and we vary the values between $\Rey_i=[1.6\times 10^2,\ 8.0\times 10^3]$. The outer cylinder is fixed. For ease of comparison with existing numerical studies, we also define the Taylor number,
\begin{equation}\label{eq:Ta}
\Tay\equiv\dfrac{(1+\eta)^6}{64\eta^4}\Rey_i^2,
\end{equation}
with the corresponding values to the $\Rey_i$ range being $\Tay=[3.9\times10^4,9.8\times10^7]$. An overview of the cases is presented in Table \ref{tab:input}. This range of $\Tay$ covers the regimes known as Taylor vortex, wavy vortex, and turbulent Taylor vortex flow \citep{Grossmann2016}. Within the chosen $\Tay$ range, the flow experiences a series of transitions. The lowest simulated $\Tay$ is chosen to lie slightly above the regime of circular Couette flow. The onset from circular Couette flow to Taylor vortex flow is estimated to occur at $\Tay \approx 1.0 \times10^4$, which is determined from the critical Reynolds number $\Rey_{i,cr}(\eta)=(1+\eta^2)/\{2\eta\alpha^2[(1-\eta)(3+\eta)]^{1/2}\}$ with $\alpha = 0.1556$ for a resting outer cylinder \citep{esser1996}. 
The transition point from Taylor vortex to wavy vortex flow has been investigated by numerous authors \citep{ahlers1983,jones1985,langford1988}.
Under current conditions it is empirically found to lie around $\Tay=3\times10^6$, by tracking the time and volume-averaged standard deviation of the vertical velocity $u_z$ as a function of $\Tay$ (see figure \ref{fig:tc}$b$). The visualisations of the aforementioned flow regimes are shown in figure \ref{fig:vel_ta2e6}.

\section{Governing equations and numerical methods}\label{sec:govEqAndNumMethod}

\begin{table}
\setlength{\tabcolsep}{0.5em}
{\renewcommand{\arraystretch}{1.2}}
	\centering
\begin{tabular}{ c c c c c  c}
$\Tay$ & $\Rey_i$ & $N_\varphi \times N_r \times N_z$   & $ 0.1d/\eta_k$ &  $ 0.2d/\eta_k$ &  $\ell/d=0.1$
\vspace{2pt}\\ 
$3.90 \times 10^4$ & $1.60\times10^2$ & $320\times128\times240$  & $1.3$ & 2.6 & $\bullet$  \\
$1.00 \times 10^5$ & $2.56\times10^2$ & $320\times128\times240$  & $1.9$ & 3.8 & $\bullet$  \\
$1.78 \times 10^5$ & $3.42\times10^2$ & $320\times128\times240$  & $2.3$ & 4.5 & $\bullet$  \\
$3.16 \times 10^5$ & $4.56\times10^2$ & $320\times128\times240$  & $2.8$ & 5.5 & $\bullet$  \\
$3.51 \times 10^5$ & $4.80\times10^2$ & $320\times128\times240$  & $2.9$ & 5.7 & $\bullet$  \\
$4.23 \times 10^5$ & $5.27\times10^2$ & $320\times128\times240$  &   -   & 6.1 &     -      \\
$5.62 \times 10^5$ & $6.08\times10^2$ & $320\times128\times240$  & $3.4$ & 6.7 & $\bullet$  \\
$7.99 \times 10^5$ & $7.24\times10^2$ & $320\times128\times240$  &  -    & 7.5 &     -      \\
$1.00 \times 10^6$ & $8.10\times10^2$ & $320\times128\times240$  & $4.1$ & 8.1 & $\bullet$  \\
$1.39 \times 10^6$ & $9.55\times10^2$ & $320\times128\times240$  &  -    & 9.0 &    -       \\
$1.91 \times 10^6$ & $1.12\times10^3$ & $320\times128\times240$  & $5.0$ & 9.9 & $\bullet$  \\
$2.68 \times 10^6$ & $1.33\times10^3$ & $320\times128\times240$  & $5.6$ & 11.1 & $\bullet$  \\
$3.80 \times 10^6$ & $1.37\times10^3$ & $360\times144\times280$  & $6.7$ & 12.4 & $\bullet$  \\
$6.00 \times 10^6$ & $2.08\times10^3$ & $360\times144\times280$  & $7.2$ & 14.3 & $\bullet$  \\
$9.52 \times 10^6$ & $2.50\times10^3$ & $480\times192\times320$  & $13.5$& 27.0 & $\bullet$  \\
$9.75 \times 10^7$ & $8.00\times10^3$ & $768\times256\times480$  &   -   & 33.9 & - \\
\end{tabular}
	\caption{Summary of simulation parameters. The first two columns denote the driving, expressed as either $\Tay$ or $\Rey_i$. The third column presents the grid resolution for $\ell/d=0.2$. The simulated cases corresponding to $\ell/d=0.1$ are indicated with $\bullet$ and have grid resolution $640\times 256 \times 480$. $0.1d/\eta_k$ and $0.2d/\eta_k$ denote the particle size to the Kolmogorov scale for $\ell/d=0.1$ and $\ell/d=0.2$, respectively.
	}
	\label{tab:input}
\end{table}

\subsection{Carrier phase}\label{sec:carrier_phase}
The velocity field is obtained by solving the incompressible Navier--Stokes equations in cylindrical coordinates. The continuity and momentum equations read (see e.g. \citeauthor{landaulifshitz} 1987)
\begin{equation}\label{eq:mass}
\frac{1}{r} \partial_r(ru_r) + \frac{1}{r}\partial_\varphi u_\varphi +\partial_z u_z=0,
\end{equation} 
and
\begin{subeqnarray}\label{eq:momentum}
\partial_t u_\varphi + (\boldsymbol{u}\cdot\boldsymbol{\nabla}) u_\varphi + \frac{u_r u_\varphi}{r}
&=& -\frac{1}{ r} \partial_{\varphi} p + \frac{1}{\Rey} \left(\boldsymbol{\Delta} u_\varphi + \frac{2}{r^2}\partial_\varphi u_r - \frac{u_\varphi}{r^2} \right)
+ f_\varphi, \\
\partial_t u_r + (\boldsymbol{u}\cdot\boldsymbol{\nabla}) u_r - \frac{u_\varphi^2}{r}
&=& -\partial_r p + \frac{1}{\Rey} \left(
\boldsymbol{\Delta}u_r -\frac{2}{r^2}\partial_\varphi u_\varphi -\frac{u_r}{r^2}
\right)+f_r, \\
\partial_t u_z + (\boldsymbol{u}\cdot\boldsymbol{\nabla}) u_z &=& -\partial_z p + \frac{1}{\Rey} \boldsymbol{\Delta}u_z + f_z,
\end{subeqnarray}
\returnthesubequation
where $(\boldsymbol{u}\cdot\boldsymbol{\nabla}) = u_r\partial_r + r^{-1}u_\varphi\partial_\varphi + u_z\partial_z$, and $\boldsymbol{\Delta} = r^{-1}\partial_r(r\partial_r) +r^{-2}\partial_{\varphi^2} + \partial_{z^2}$. The last terms ($f_\varphi$, $f_r$, $f_z$) on the r.h.s. of \eqref{eq:momentum} denote the IBM forcing (see Appendix \ref{app:numMethod} for details).

We employ a fractional step method to numerically solve \eqref{eq:mass} and \eqref{eq:momentum}. The velocity field is discretised using a conservative spatial, second-order, central finite-difference scheme and a temporal third-order Runge--Kutta scheme, except for the viscous terms that are treated implicitly with a Crank--Nicolson scheme. In the wall-normal direction, the grid is stretched with a clipped Chebychev type of stretching \citep[\cf][]{ostilla2015}. The grid is uniform in the azimuthal and axial directions. For more details, we refer the reader to \cite{Verzicco1996}.
The grid resolution for the fluid phase is based on  \cite{ostilla2013}, with the note that the grid aspect ratio here is 1.0 at mid-gap. This criterion is used to ensure sufficient nodes for the IBM, which is discussed in \S\,\ref{sec:dispersed_phase}. The time-step is variable and satisfies the condition $\text{CFL}=0.3$; this restrictive limit is used owing to the explicit coupling of the particles to the fluid phase.

\subsection{Particles} \label{sec:dispersed_phase}

We use prolate ellipsoids as the dispersed phase in the TC flow. The control parameter for the particle is the ratio of particle size to the gap width, $\ell/d$, with $\ell$ the major axis of the ellipsoid (figure \ref{fig:tc}$a$). For our study $\ell/d=0.1$ or $0.2$. The aspect ratio of the ellipsoid is $\Lambda\equiv\ell/b= 4$, with $b$ the minor axis of the particle. 16 particles are used in each simulation, yielding volume fractions of $0.01\%$ and $0.07\%$, for $\ell/d=0.1$ and $0.2$, respectively. The reported Stokes number is obtained via \citep[\cf][]{Voth2017}

\begin{equation}\label{eq:stokes}
St\equiv\frac{\tau_p}{\tau_v},\quad \text{with}\quad \tau_p=
 \frac{1}{18}\frac{b^{2}}{\nu}\frac{\Lambda\ln({\Lambda + \sqrt{\Lambda^2-1}})}{\sqrt{\Lambda^2-1}}\quad \text{and}\quad \tau_v=\dfrac{\nu}{u^2_\tau}.
 \end{equation}
Here, the friction velocity is $u_\tau=\sqrt{\nu\partial_r \langle u_{\varphi}\rangle_{A,t}}|_{r_i}$ (average over time and inner cylinder).

The rigid particle dynamics are obtained by integrating the Newton--Euler equations. The governing equations and numerical methods regarding the particle translation, rotation, and collisions are provided in Appendix \ref{app:numMethod}.

The ratio of the particle size to the Kolmogorov scale is estimated based on the global exact balance for TC flow \citep{eckhardt2007} 
 \begin{equation}\label{eq:kolmogorov}
 \eta_K/d = \left(\sigma^{-2} \Tay [\Nus_\omega-1] \right)^{-1/4},\quad \text{where}\quad\sigma = (1+\eta)^4/(16\eta^2),
 \end{equation}
 and $\Nus_\omega = J^\omega/J^\omega_{\text{lam}}$, the Nusselt number. Here, $J^\omega = r^3(\langle u_r \omega\rangle_{A,t} - \nu \partial_r \langle \omega \rangle_{A,t})$ and $J^\omega_{\text{lam}}= 2\nu\omega_i r_i^2r_o^2 /d^2 $. 
 For all $\Tay$, we estimate $\ell/\eta_k$ based on \eqref{eq:kolmogorov}, see Table \ref{tab:input}. 

To initialise the simulations, the single-phase flow is first advanced in time until a statistically stationary state is attained. Once this state has been achieved, the particles are inserted at random positions, with zero velocity within the domain, whilst ensuring that their initial distribution is spatially homogeneous. The initial orientations of the particles are also randomised. After inserting the particles, the simulations are run for at least 50 flow-through times before collecting statistics. A number between 15 to 25 grid points per $\ell$ are used to ensure the boundary layers over the particles be sufficiently resolved. The particle Reynolds number is  estimated as  $\Rey_p\equiv \dot{\gamma} \ell^2 /\nu$ and ranges from O(0.1) to O(60), with $\dot{\gamma}$  the average radial derivative of the azimuthal velocity in the bulk.

\section{Spatial distributions of particles}\label{sec:SpatialDistributions}
\subsection{Observed spatial modes}\label{sec:spatial_distributions}
\begin{figure}
\includegraphics[width=\linewidth]{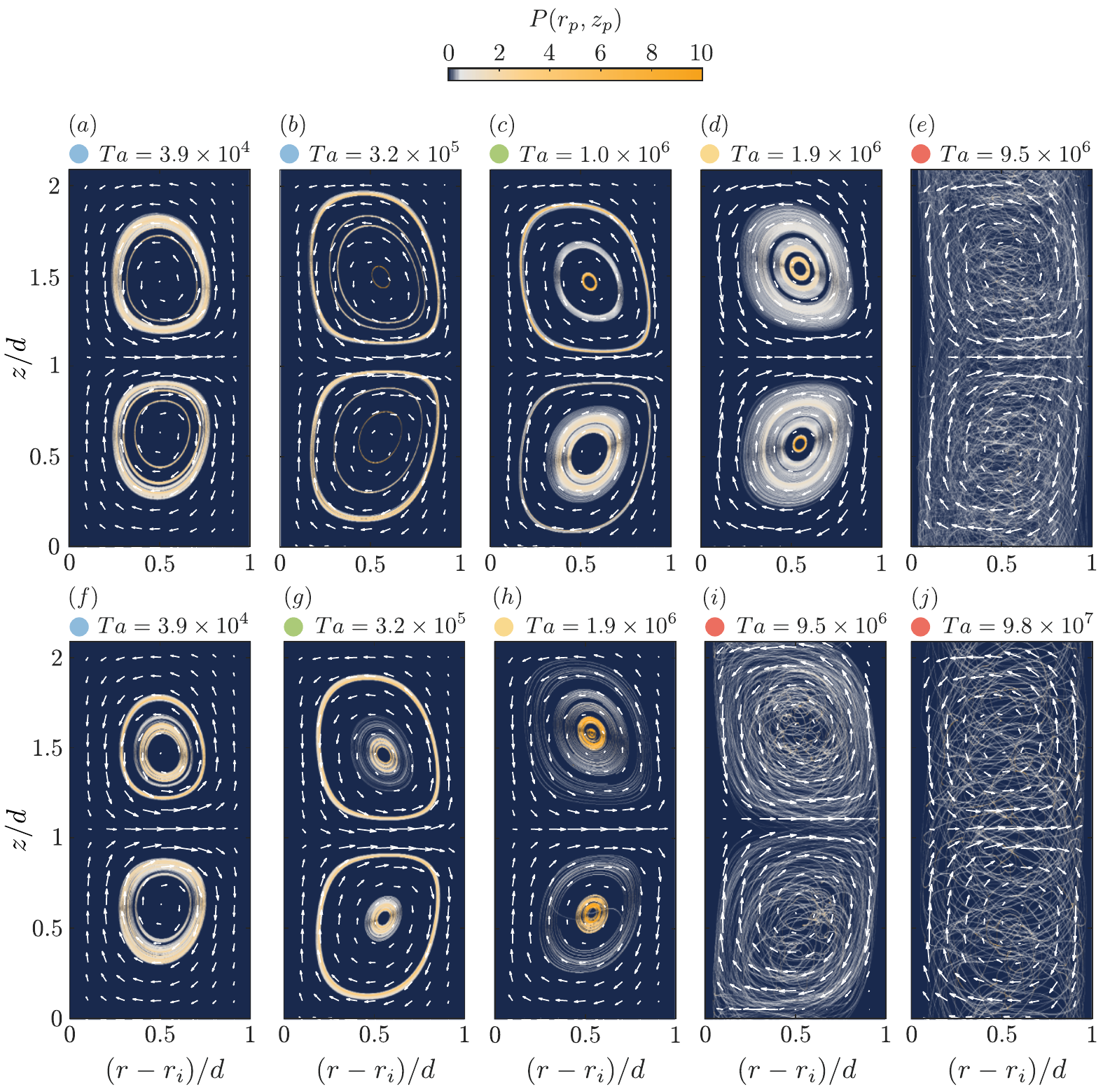}
\caption{Probability density function of the particle distribution. The average is taken over time and azimuthal direction. $(a)-(e)$ correspond to particles of size $\ell/d=0.1$ and $(f)-(j)$ to particles of $\ell/d=0.2$. The coloured circles on top of the contour plots denote distinguishable regimes of particle dynamics, which correspond to those in figure \ref{fig:pdf_rad_pos} and \ref{fig:orientations}.}
\label{fig:position_r_z}
\end{figure}

We examine the statistics of the particle positions. In particular, we select the cases $\Tay=3.9\times10^4,\ 3.2\times 10^5,\ 1.9\times10^6,\ 9.5\times10^6$ and $9.8\times10^7$ with particles (see also the single phase flow fields in figure \ref{fig:vel_ta2e6}). In figure \ref{fig:position_r_z} we present the time-averaged particle distribution, after the initial transients, projected onto the $r-z$ plane for both particle sizes; $\ell/d=0.1$ and $\ell/d=0.2$.  Figure \ref{fig:position_r_z} reveals four distinct spatial patterns, which depend on both $\ell/d$ and $\Tay$. The characteristics of these flow different "modes" are as follows: 
\medskip
\begin{enumerate}[label= Mode \itshape{(\roman{*}): }, align =left]
 	\item Steady large orbits (figure \ref{fig:position_r_z}$a,b,f$).
 	\item Steady orbits, with particles spiralling closely around the vortex cores (figure \ref{fig:position_r_z}$d,h$).
 	\item A combination of modes $(i)$ and $(ii)$ (figure \ref{fig:position_r_z}$c,g$).
 	\item Unsteady orbits, with particles distributed quite homogeneously throughout the domain (figure \ref{fig:position_r_z}$e,i,j$).
 \end{enumerate}
\medskip
For mode $(i)$, the rotational particle dynamics show no stable alignment, but instead, a tumbling type of motion is observed. At this $\Tay$, the base flow is slightly above the transition point from the circular Couette flow  to a Taylor vortex flow regime. We stress that the particles were released at random locations after the flow was fully developed -- therefore, on release, each particle undergoes an inertial migration process before reaching its  stable orbit. 

For mode $(ii)$, particle orbits are observed to agglomerate at the vortex cores. The most pronounced example from figure \ref{fig:position_r_z} is  at $\Tay=1.9\times 10^6$ for $\ell/d=0.2$ (panel {\it h\/}). Remarkably, the particle concentration at the core is much higher for the larger particles, thus indicating that this is definitely a finite-particle-size effect. Mode $(ii)$ is accompanied also by a stable particle alignment, which will be addressed in detail in \S\,\ref{sec:AngularStatistics}.

Mode $(iii)$ consists of a combination of modes $(i)$ and $(ii)$. This regime is the transition between stable (limit cycle-type) orbits and preferential clustering at the core. This mode is observed at $\Tay=1.0\times10^6$ for $\ell/d=0.1$. Interestingly, for $\ell/d=0.2$, mode $(iii)$ is observed at a lower value of $\Tay=3.2\times 10^5$. Hence, it appears that both particle clustering as well as the transitions to various particle orbit regimes are functions of $\ell/d$. In \S\,\ref{sec:pdf_r}, we will study the regime transitions as a function of $\ell/d$.

One key feature of modes $(i)$ up to $(iii)$ is the steadiness of the orbits as the particles trace their path through the domain. This steadiness occurs at a unique set of conditions and can be linked to two fundamental features of the Taylor vortices. Firstly, the background flow is completely steady and time-invariant. Secondly, the Taylor vortices are axisymmetric about the cylindrical axis (see \S\,\ref{sec:Setup}). 

For mode $(iv)$, we now observe unsteady dynamics caused by unsteady Taylor vortices. For instance, as shown in figure \ref{fig:vel_ta2e6}($e$), the flow corresponds to the wavy Taylor vortex regime. Due to the unsteadiness of the wavy Taylor vortices, particles experience spatial and temporal variations of the hydrodynamic loads. These variations prevent any stable orbits from happening. 
For $\Tay=9.5\times10^6$ and $\ell/d=0.2$, particle trajectories tend to maintain some coherence (see figure \ref{fig:position_r_z}$i$) and appear to trace out the complete vortex. However, for the same $\Tay$ and $\ell/d=0.1$ (figure \ref{fig:position_r_z}$e$), as well as at larger $\Tay$ for $\ell/d=0.2$ (figure \ref{fig:position_r_z}$i$), the coherence observed for mode ($ii$) is  lost and the particles distribute nearly homogeneously  throughout the domain. The distributions for the latter combinations of $\Tay$ and $\ell/d$ are reminiscent of those for spheres and fibres in TC \citep{Majji2018,Bakhuis2019}.


\subsection{The transition from stable orbits to clustering at the core is size-dependent}\label{sec:pdf_r}
\begin{figure}
	\centering
 \includegraphics[width=\linewidth]{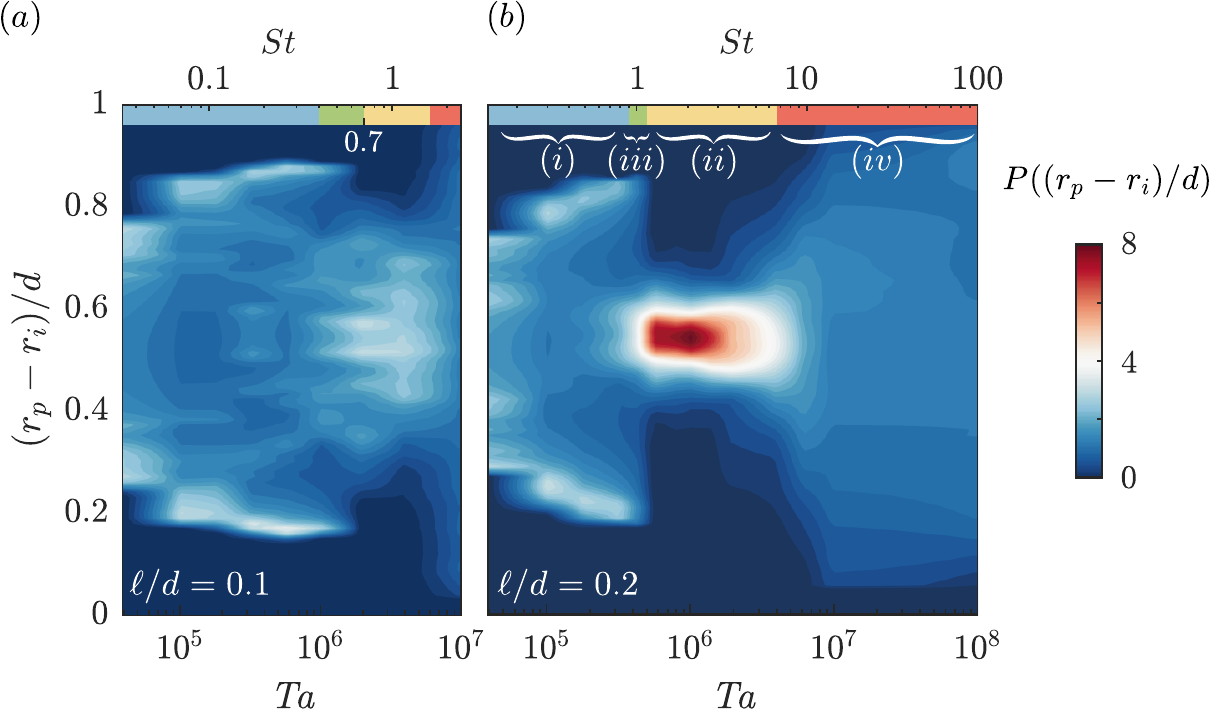}
	\caption{Joint pdf of the particles normalised radial position versus $\Tay$. Particles with size ratio {\it (a)} $\ell/d=0.1$. {\it (b)} $\ell/d=0.2$. For reference, the modes addressed in \S\,\ref{sec:spatial_distributions} are indicated with $(i)$-$(iv)$. The coloured top bar uses colour codes corresponding with those in figure \ref{fig:position_r_z} and \ref{fig:orientations}.}
	\label{fig:pdf_rad_pos}
\end{figure}
In \S\,\ref{sec:spatial_distributions},  in some cases particles are observed to preferentially cluster at the vortex core. 
Now, we will examine the clustering behaviour in more detail. 
The $\Tay$ range for which clustering occurs is investigated via the pdf of the particle radial distributions $P(r_p)$, with $r_p$ being the particle radial position. $P(r_p)$, is plotted versus $\Tay$ in figure \ref{fig:pdf_rad_pos}. From this figure, two main observations can be made. (1) We observe that the modes, indicated by the colours at the top of the figures, do not  occur  exactly at the same $\Tay$ when comparing $\ell/d=0.1$ and $\ell/d=0.2$. (2) The magnitude of the peak is much more intense for  $\ell/d=0.2$, suggesting that clustering is enhanced for the larger particles. These two effects are discussed below.

In \S\,\ref{sec:spatial_distributions} we defined four modes characterised by the particle distributions in the flow field.
We highlighted these regimes in figure \ref{fig:position_r_z} with four colours at the top; the colours correspond to those used in figure \ref{fig:pdf_rad_pos}. Mode $(i)$ corresponds to the stable particle orbits, resulting in helical particle trajectories. For this regime, we observe a preferential particle concentration away from the vortex centre, as is evident by the light blue regions at $(r_p-r_i)/d$ around $ 0.2$ and $ 0.8$ in figure \ref{fig:pdf_rad_pos}. In this regime, particles  in the Taylor-vortex are found to move further outwards. Beyond this regime for even larger $\Tay$, we observe mode $(iii)$, which is the mixed regime in which both modes $(i)$ and $(ii)$ are observed simultaneously. Beyond this, mode $(ii)$ is encountered; particles cluster at the central region of the Taylor vortices as evidenced by the peak in the pdf at $(r_p-r_i)/d \approx 0.5$. Finally, beyond this regime, particles move outwards again and distribute more homogeneously when the Taylor vortex starts to destabilise into the wavy regime. When comparing figures \ref{fig:pdf_rad_pos}($a$) and ($b$), the transition  between the regimes shifts to lower $\Tay$ for larger particles. Since there is this clear size dependence, it is, therefore, instructive to compare the corresponding particle time scale $\tau_p$, which is set by the particle size, with the fluid shear time scales, $\tau_\nu$: effectively, we compute the Stokes number $\Stk$ for the particles as defined earlier in \eqref{eq:stokes}. 
When considering $\Stk$ as the governing parameter, we observe that mode $(iii)$ (transition to clustering) occurs at $\Stk \approx 0.7$ for $\ell/d=0.1$ and $\Stk\approx 1.0$ for $\ell/d=0.2$, suggesting that $\Stk$ is of a similar order of magnitude for mode ($iii$).
However, the reason why we urge caution is that, aside from the  small numerical parameter space, TC flows are inherently three-dimensional because of curvature effects. Therefore, particle dynamics become sensitive to spatially varying hydrodynamic forces \citep{trevelyan1951}. 
Curvature effects on particles in TC can be straightforwardly investigated by examining Jeffery solutions \citep{Jeffery1922} for prolate ellipsoids in the Couette regime. In appendix \ref{app:curv_effects} evidence is provided that curvature effects on the particle rotation rate are already observed at Taylor numbers as low as $\Tay=1.0$.

\medskip
Comparing clustering for the two different particle sizes, we observe a much higher magnitude of $P(r_p)$ in figure \ref{fig:pdf_rad_pos} for $\ell/d=0.2$ than for the case $\ell/d=0.1$. The clustering is weaker for smaller $\ell/d$ for the following reason: the clustering regime of $\ell/d=0.1$ falls together with the onset of the wavy Taylor vortex regime, while particles of $\ell/d=0.2$ start to cluster around $\Tay\approx4\times 10^5$ (Taylor vortex regime).

\section{Statistics of the particle orientation}\label{sec:AngularStatistics}
\subsection{Angular dynamics}\label{sec:orientationDefinition}
\begin{figure}
\includegraphics{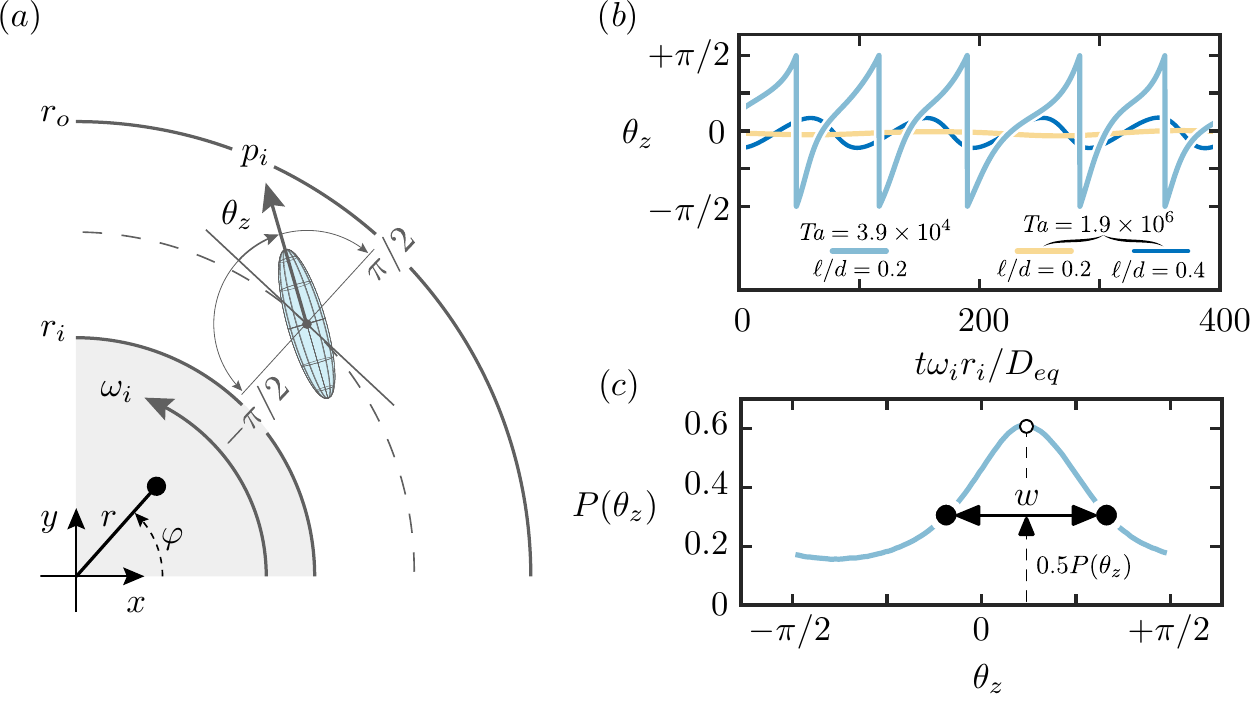}
\caption{({\it a\/}) Definition of the angle $\theta_z$ between the pointing vector $p_i$ and the tangent along the cylinder. By symmetry of the particle $\theta_z \in [-\pi/2,\pi/2]$. ({\it b\/}) Angular time signal of a particle within a stable orbit (light blue line, $\Tay=3.9\times10^4$ \cf figure \ref{fig:position_r_z}$(f)$ and  for mode ($ii$)  (yellow line, $\Tay=1.9\times 10^6$ \cf figure \ref{fig:position_r_z}$h$). ({\it c\/}) Definition of the width of the pdf $P(\theta_z)$. The width is measured for the highest peak of $P(\theta_z)$ at half-height. }
\label{fig:angle}
\end{figure}

Up to this point, the  spatial statistics of particles have been examined. A number of regimes was found, one of which is of specific interest since particles were found to cluster at the central region of Taylor vortex cores. Additionally, clustering is observed to enhance when the particle size is increased from $\ell/d=0.1$  to $\ell/d=0.2$. As a follow-up, we examine the statistics of particle orientations, corresponding to the identified spatial modes. We examine the angle
 $\theta_z$ (see figure \ref{fig:angle}$a$) between the particle pointing vector $p_i$ and the local tangent of the cylinder \citep[\cf][]{Bakhuis2019}. Here, we make use of the symmetry of the particle and let $\theta_z \in [-\pi/2,\pi/2]$ .

Two typical time signals of $\theta_z$ are given in figure \ref{fig:angle}($b$) for particles of size $\ell/d=0.2$. The signal for $\Tay=3.9\times10^4$ belongs to a particle travelling along a stable orbit. Interestingly, the particle orientational dynamics in the steady Taylor vortex regime still show a Jeffery-like character. These Jeffery-like features have been observed in a variety of cases that also do not strictly fulfil the criterions of Jeffery's equations \citep{wang2018,kamal2020}. 
In contrast, the time signal of $\theta_z$ for $\Tay=1.9\times 10^6$ shows an interesting, nearly-constant angle $\theta_z$. This may be visualised as if the axis of revolution always aligns with the local cylinder tangent. The particle does not flip but oscillates only relatively mildly. This case corresponds to preferential particle concentration at the vortex core (mode $(ii)$, see figure \ref{fig:position_r_z}$h$). 

For illustration purposes eight particles of size $\ell/d=0.4$ at  $\Tay=1.9\times10^6$ are simulated (subject to same grid resolution for the corresponding case $\ell/d=0.2$). This larger particle displays mode $(ii)$, too, with a slightly larger oscillatory character (see figure \ref{fig:angle}$b$). This indicates that finite-size effects remain visible for even larger particles, but non-linear effects come into play, which are out of the scope of this study. 

The statistics of $\theta_z$ are discussed in the following and linked to the spatial distribution regimes of \S\,\ref{sec:spatial_distributions}. 

\subsection{Angular statistics corresponding to the observed spatial distributions} \label{subsec:AngularStatistics}
\begin{figure}
\includegraphics[width=\linewidth]{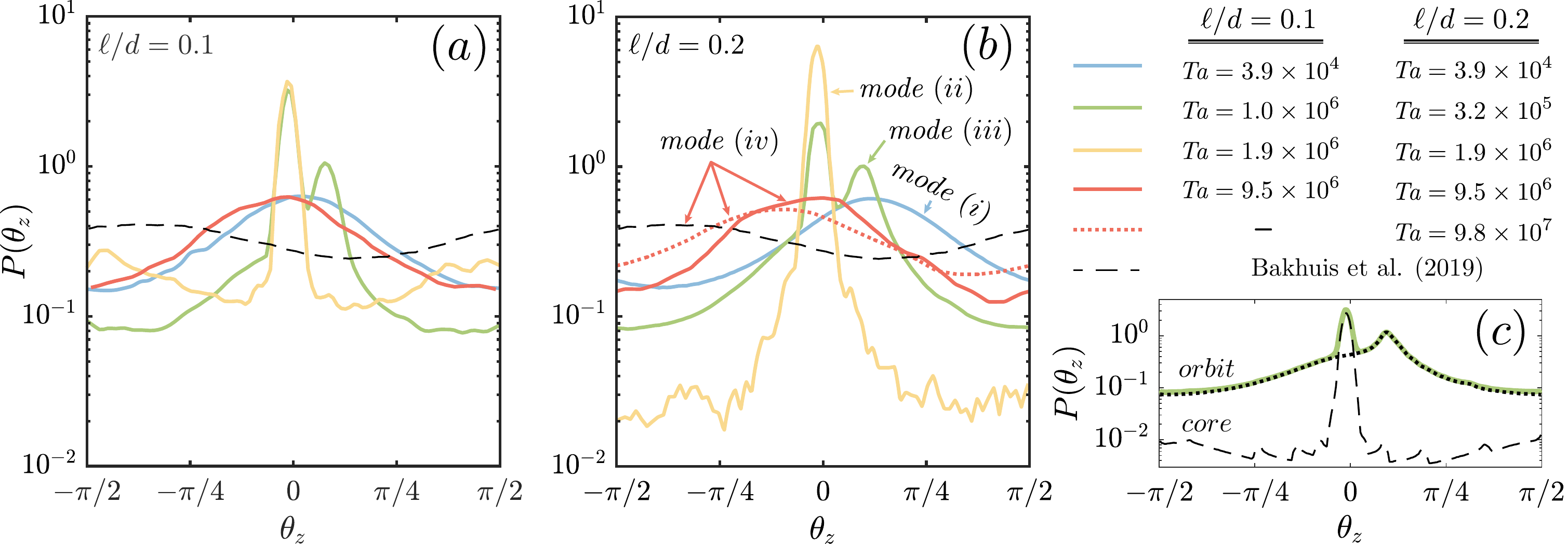}
\caption{The pdf of $\theta_z$ for various $\Tay$. ({\it a\/}) Particles with size $\ell/d=0.1$, ({\it b\/}) Particles with size $\ell/d=0.2$, ($c$) Decomposition of the orientational statistics for mode $(iii)$, showing the origin of the two peaks. A fraction of the particles is close to the vortex core, whereas other particles are within a stable orbit. For reference, the experimental observations from \citet{Bakhuis2019} are added (\cf $\Tay=9.5\times10^{10}$ ). The colour coding of the plots is in correspondence with figures \ref{fig:position_r_z} and \ref{fig:pdf_rad_pos}. }
\label{fig:orientations}
\end{figure}

Typical pdfs of $\theta_z$ for various $\Tay$ and for $\ell/d=0.1$ and $\ell/d=0.2$  are given in figure \ref{fig:orientations}($a,b$). The shown cases correspond to the spatial distributions in figure \ref{fig:position_r_z}. 
Several interesting features can be observed in the distribution of $P(\theta_z)$. In particular, we find that these features correlate to the different  spatial particle distributions  described earlier in \S\,\ref{sec:spatial_distributions} as listed  below.
\medskip
\begin{enumerate}[label=\itshape{(\alph{*})}, leftmargin=0.5cm,labelsep=0.2cm]
 	\item Steady large orbits. - A positive preferred orientation (maximum of $P(\theta_z)$ occurs at $\theta_z>0$).
 	\item Orbits spiralling closely around the core. - A sharp peak for $P(\theta_z)$ located at $\theta_z\approx 0$.
 	\item A combination of modes ($i$) and ($ii$). - Angular dynamics show modes ($i$) and ($ii$) on the border between clustering and stationary orbits.
 	\item Unsteady dynamics,  particles distribute throughout the whole domain. - A non-homogeneous distribution of $\theta_z$ with the maximum of $P(\theta_z)$ occurring at negative $\theta_z$.
 \end{enumerate}
 \medskip
For mode ($i$), the flow is in the steady Taylor vortex flow regime ($\Tay=3.9\times 10^4$). From figure \ref{fig:orientations}, for both $\ell/d=0.1$ and $\ell/d=0.2$ the angular statistics display  a non-homogeneous distribution, with a positive preferred angle. 

For mode ($ii$), particles agglomerate at the Taylor vortex cores. Remarkably they show a strong preferential alignment at $\Tay=1.9\times10^6$ as confirmed by the sharp peaks of the alignment probability of  figure \ref{fig:orientations}. The more defined peak of $P(\theta_z)$ observed for $\ell/d=0.2$, as compared to $\ell/d=0.1$, is related to the enhanced clustering (see \S\,\ref{sec:pdf_r}) and correlates with the result of stronger preferential alignment, thus indicating that the particle size plays a predominant role in the phenomenon. Tails of $P(\theta_z)$ can also be observed, for example at $\Tay=1.9\times 10^6$ in figures \ref{fig:orientations}($a,b$). These tails are the result of particles precessing in a stable orbit visible in the particle distribution in figures \ref{fig:position_r_z}($d,h$). Our explanation why particles may still intermittently exhibit behaviour akin to ``stable orbits'' is because particles collide throughout these simulations at this $\Tay$ (O(5) for all particles per flow-through time). Observations from such events indicate that collisions cause the particles to end up further away from the vortex core in a meta-stable orbit, which eventually decays to the stable preferential alignment at the vortex core.

For mode ($iii$), two peaks are observed for $P(\theta_z)$, as shown by the green curves in figures \ref{fig:orientations}($a,b$). These two peaks originate from the two spatial modes ($i$) and ($ii$), shown in figure \ref{fig:position_r_z}($g$). The contribution from the two modes can be explained by disentangling the two angular statistics, which we illustrate in figure \ref{fig:orientations}($c$): First, we separate the two spatial modes by taking a sub-sample of particles close to and far from the vortex cores. Next, the angular statistics of these sub-samples are computed, resulting in two pdfs of $\theta_z$ (black dashed lines in figure \ref{fig:orientations}$c$). These separate pdfs illustrate that particles close to the vortex core show preferential alignment at $\theta_z\approx 0$, whereas those in a stable orbit far from the core peaks at $\theta_z\approx 0.09\pi$. It is highlighted that this particle behaviour forms the transition point between stable orbits and clustering (mode ($iii$) in figure \ref{fig:pdf_rad_pos}), and occurs within a very narrow $\Tay$ range. 

For mode ($iv$), particles distribute homogeneously throughout the domain due to the instabilities and unsteadiness of the flow. The preferential alignment observed with axisymmetric Taylor vortices (mode $ii$) cannot exist when the flow undergoes a transition to wavy Taylor vortex flow. For $P(\theta_z)$ this results in a distribution that is flatter. However, some statistical preferential alignment persists. \citet{Bakhuis2019} reported a difference of 40\% between the lowest and highest values of $P(\theta_z)$ at $\Tay=1.0\times10^{12}$ for cylinders of aspect ratio $\Lambda=5$. Within this work, at $\Tay = 9.8 \times 10^8$, the difference is about 67\% which suggests that the tendencies for particles to preferentially align are stronger at lower $\Tay$. We also highlight that, while the $\Tay$ values are much lower in our setup than in \citet{Bakhuis2019}, there is a general tendency for the peaks to shift for  increasing $\Tay$ towards negative $\theta_z$ and lower $P(\theta_z)$. Indeed, the incipient trend is consistent with the distributions in \citet{Bakhuis2019}. Further studies at larger $\Tay$ in the simulations will be necessary to verify this trend.

\subsection{The most pronounced alignment of particles}\label{sec:mostPronouncedAlignment}
\begin{figure}
\centering
\includegraphics[width=0.75\linewidth]{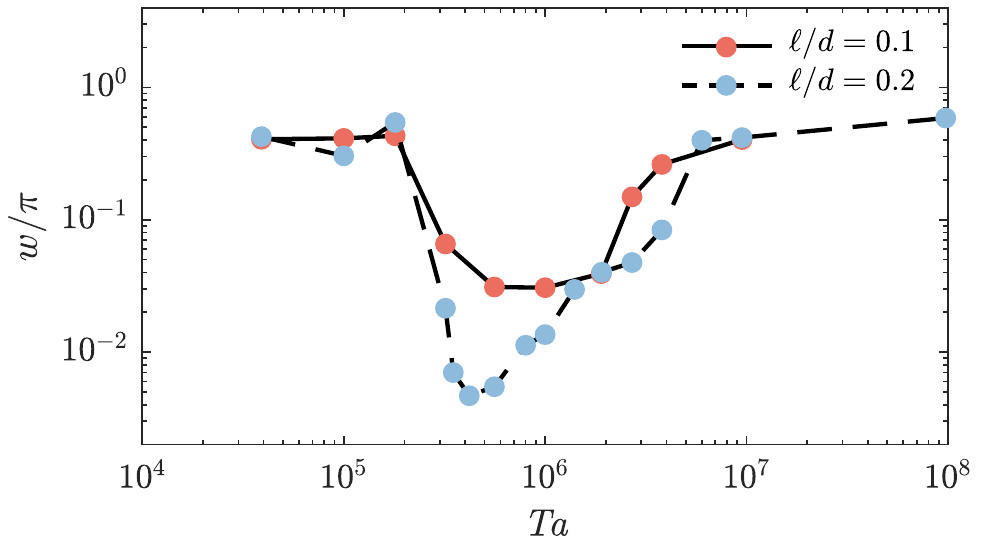}
\caption{Width, $w$, of the pdf $P(\theta_z)$ versus $\Tay$. The definition of $w$ is sketched in figure \ref{fig:angle}}
\label{fig:widthVsTa}
\end{figure}

In \S\,\ref{subsec:AngularStatistics}, preferential alignment of the particle angle $\theta_z$ is observed in the case when particles agglomerate near the vortex core. Here, the objective is to determine the conditions for which alignment is strongest. 
For all $P(\theta_z)$, the width  $w$ of the pdf around the highest peak is calculated and taken at half-height (see figure \ref{fig:angle}$c$). 
The plot of $w$ versus $\Tay$ is given in figure \ref{fig:widthVsTa}.
Remarkably, $w$ has a very pronounced minimum (note the double log-scale) with respect to $\Tay$, occurring around $\Tay=7\times 10^5$ for $\ell/d=0.1$ and at $\Tay=4\times10^5$ for $\ell/d=0.2$. This minimum corresponds to a particle that oscillates the least with respect to the local tangent of the cylinder. 
Intriguingly, the minimum is even more pronounced for $\ell/d=0.2$ compared to $\ell/d=0.1$. In the remainder of this work, we will discuss the origin of this minimum in $w$ and offer an explanation for why this is observed in the investigated configuration.

\section{The role of axial vorticity at the vortex core.}\label{sec:optimality explanation}
\subsection{The link between strong alignment and minimum axial vorticity }
From our analysis in the preceding sections, we observed that at specific $\Tay$ values particles tend to get trapped within the vortex core  and have a preferred orientation. Now, we aim to answer the question: Why do the particles align at this $\Tay$ value? In the following, we will show that the preferential alignment is linked to the TC flow state which exhibits a minimum in the shear gradient at the Taylor vortex core. In particular, we will base our analysis on the axial component of the vorticity of the flow, which is shown to be the key metric determining the preferential alignment.

\begin{figure}
	\centering
	\includegraphics[width=0.95\linewidth]{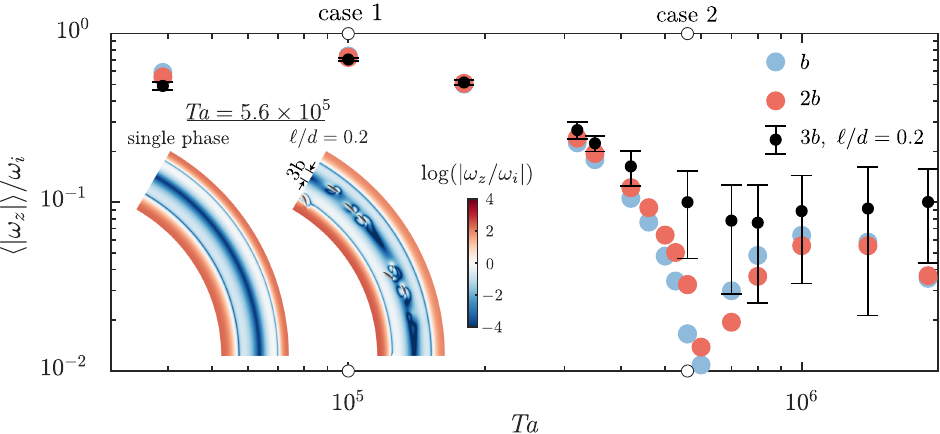}
	\caption{The average vorticity, $\langle | \omega_z | \rangle/\omega_i$, is computed for the single phase flow situation. The area covered for the average vorticity, $\omega_z$, is a circle centred at the vortex core with radius $b$ and $2b$, where $b$ denotes the particle minor axis with $\ell/d=0.2$. The analysis for a patch with radius $3b$ is performed with the presence of particles of size $\ell/d=0.2$.
	The inset shows an instantaneous $\varphi$-$r$ slice of $\omega_z$ for the single-phase flow and two-phase flow cases, highlighting the perturbed vorticity fields due to the presence of particles. }
	\label{fig:avevort}
\end{figure}

In the spirit of Jeffery's equation for the rotation of an ellipsoid, we formulate an area average of the axial vorticity, $\omega_z=r^{-1} \partial_{r} ( r u_\varphi) - r^{-1}\partial_\varphi u_r$ evaluated in the $r,z$ plane, based on the premise that a particle aligns with the local cylinder tangent. This vorticity component is held
responsible for rotational flipping events of the ellipsoidal particle around the $z$-axis. To compute the axial vorticity, it is first instructive to identify a region-of-interest where particles would presumably cluster. Taking heed from the clusterings observed in figure \ref{fig:position_r_z}, this region can be reasonably and safely assumed to coincide with the central regions of the Taylor vortices. Therefore, the average of $\omega_z$ is taken for each $r-z$ plane over a circular patch positioned at the vortex cores. The patch has a radius assigned with cross-section dimensions similar to that of the particle. Here, we select a circle with radius $i\cdot b$, with $b$ the minor axis of the particle. The Taylor vortex cores are identified by a local minimum of the meridional velocity, $u_r^2+u_z^2$. Note that for the single-phase flow in the Taylor vortex regime, the vorticity in the $r,z$ plane is stationary and, due to the axial symmetry of the flow, independent of the coordinate $\varphi$. The averaged $\omega_z$  is normalised by the rotational velocity of the inner cylinder $\omega_i$ and plotted in figure \ref{fig:avevort} versus $\Tay$.

A minimum of $\langle |\omega_z|\rangle$ can be clearly observed at $\Tay \approx 6\times10^5 $. This minimum implies that, roughly close to this $\Tay$ value, the fluid torque applied to the body (following Jeffery's equations) will be the smallest. Indeed, comparing this minimum to the previously acquired most pronounced alignment in figure \ref{fig:widthVsTa}, a very good agreement is found: The preferential alignment of the particle at $\Tay=4.2\times10^5$ and the minimum of average vorticity for the single-phase flow occurs at $\Tay=6\times10^5$. Additional calculations with a larger nominal diameter equal to $2b$ (red circles in figure \ref{fig:avevort}) also find the minimum to occur around this $\Tay$, which shows that this metric is quite robust. 

For completeness, we further analyse the effect on $\langle|\omega_z|\rangle$ with particles for $\ell/d=0.2$. In contrast to the single-phase flow, the presence of particles introduces velocity gradients in their vicinity because of the formation of viscous boundary layers at the particle surface. Therefore, $\langle |\omega_z|\rangle$ is determined at a slightly larger patch with radius $3b$ in order to filter out spurious vorticity magnitudes with particles. Here, the volume occupied by the particle is excluded from the calculation. Patches smaller than $3b$ are possible although they result in larger variances due to the closer proximity to the wakes shed by the particles. This trend of $\langle|\omega_z|\rangle$ including the particles is shown in figure \ref{fig:avevort} with black symbols. As can be seen from the figure, the trend is closely followed, but when particles agglomerate near the vortex core, the metric becomes very sensitive to particle-induced gradients and skews the picture. We find the lower values of $\langle|\omega_z|\rangle$ to occur in parts of the domain in which the particle is absent. The pronounced results of $\langle|\omega_z|\rangle$ for the single-phase flow served as a good guideline in our understanding of strong alignment. However, distilling similar results with particles is challenging. 

\subsection{Lagrangian statistics of a particle rotational energy}\label{subsub:lagrangiankineticenergy}

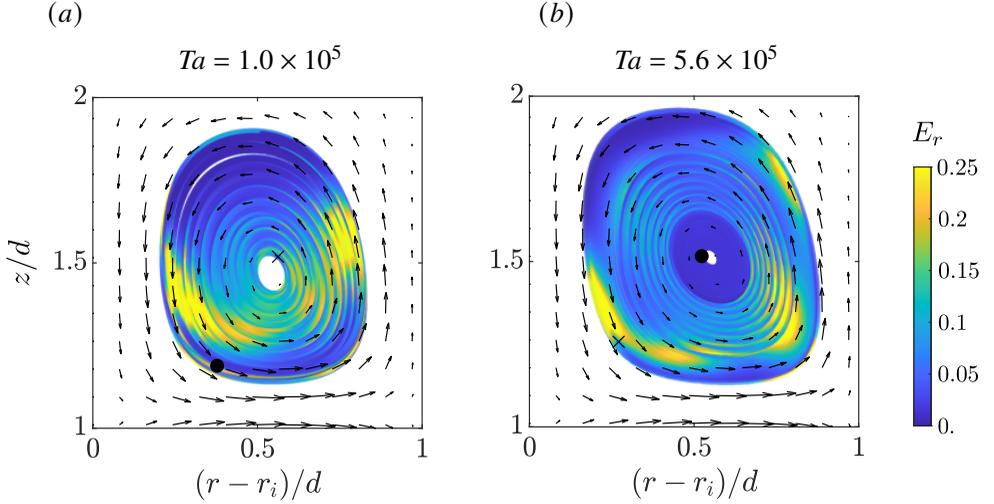
\begin{figure}
\centering
\begin{minipage}{0.49\textwidth}
\centering
	\begin{tikzpicture}
	\tikzstyle{every node}=[font=\normalsize\selectfont]
	\node[anchor=south west,inner sep=0] (image) at (0.0,0) {\includegraphics[height = 0.85\textwidth]{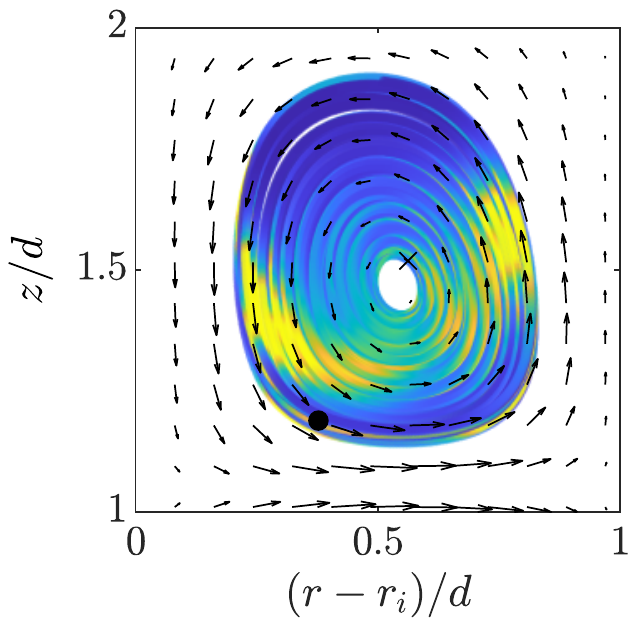}};
	\begin{scope}[x={(image.south east)},y={(image.north west)}]
	\node [above] at (0.15,1.1) {$(a)$};
	\node [above] at (0.6,1.0) {$\Tay=1.0\times10^5$};
	\end{scope}
	\end{tikzpicture}
\end{minipage}
\hfill
\begin{minipage}{0.49\textwidth}
\centering
		\begin{tikzpicture}
	\tikzstyle{every node}=[font=\normalsize\selectfont]
	\node[anchor=south west,inner sep=0] (image) at (0,0) {\includegraphics[height=0.85\textwidth]{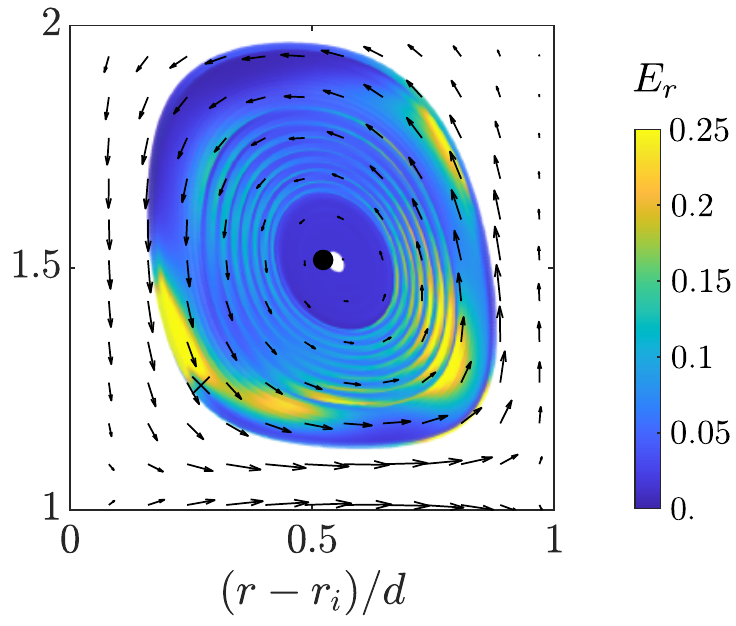}};
	\begin{scope}[x={(image.south east)},y={(image.north west)}]
	\node [above] at (0.15,1.1) {$(b)$};
	\node [above] at (0.43,1.0) {$\Tay=5.6\times10^5$};
	\end{scope}
	\end{tikzpicture}
\end{minipage}
\vspace{4pt}
	\caption{The dimensionless space-time evolution of the rotational energy, $E_r$, of a particle for $(a)$ $\Tay=1.0\times10^5$ and $(b)$ $\Tay=5.6\times10^5$. In $(a)$, the particle eventually spirals outwards and does not display mode ($ii$). In $(b)$, the particle spirals inwards towards the core. The starting and ending positions of the particle are denoted by $\times$ and $\sbullet[1.5]$, respectively.
	Large magnitudes of $E_r$ correspond to tumbling events of the particle. The arrows denote the $(u_r,u_z)$ velocity field.}
	\label{fig:rotkinenta56e5}
\end{figure}
In this final analysis we investigate the effect of $\omega_z$ on the particle dynamics in relation to the particle position. Here, we select a single representative  particle ($\ell/d=0.2$) from cases $\Tay=1.0\times10^5$ and $\Tay=5.6\times10^5$. For the purpose of our discussion we refer to these as case 1 and 2, respectively. Note that both cases are highlighted with a circle at the top of figure \ref{fig:avevort}. Case 1 corresponds to a particle exhibiting a final spatial distribution with a stable orbit far from the vortex core, whereas case 2 converges to a strong alignment mode in the vicinity of the vortex core. For both cases the rotational kinetic energy is tracked over time (translational energy is left out of the analysis). Here, because of the tumbling motion of the particle, it is assumed that its rotational energy provides a reasonable metric of the particle Lagrangian dynamics. The particle selected for case 1 initially started out close to the vortex core, whereas case 2 initially started at the edge of the vortex and subsequently  spiralled inwards.

The particle rotational energy, $E_r$, is given by 
\begin{equation}
E_r = \frac{1}{2}\hat{\boldsymbol{\omega}}^T \boldsymbol{I}_p \hat{\boldsymbol{\omega}},
\end{equation} 
with $\hat{\boldsymbol{\omega}}=\boldsymbol{\omega}_p/\omega_i$  the normalised rotational velocity and $\boldsymbol{I}_p$ its moment of inertia tensor of the particle. The two cases are illustrated in figure \ref{fig:rotkinenta56e5}. We observe local regions where the rotational kinetic energy is highest, which correspond to particle tumbling events. The distinct difference between case 1 and 2 is when $E_r$ is examined at the vortex core. In fact, for case 2, the value of $E_r$ at the core is negligibly small and is well-correlated with the low $\langle|\omega_z| \rangle$ value shown in figure \ref{fig:avevort}. This local minima, interestingly, only holds for a specific region close to the core, whereas outside of the core, the rotational kinetic energy is finite. This picture, therefore, illustrates that strong alignment occurs only when particles are within the local vorticity minima of the core. In contrast, case 1 clearly shows tumbling events that can still be found at the vortex core. This phenomenon, therefore, illustrates the importance of axial vorticity in determining the preferential alignment of ellipsoids in TC  flow.

Based on our findings on the vorticity statistics, this secondary mechanism appears also to be a crucial factor that determines preferential alignment, in addition to the Stokes number effect discussed in \S\,\ref{sec:pdf_r}. Simply put, while clustering can be controlled by varying $\Stk$ (or $\ell/d$), the unique non-linear axial vorticity fields at different $\Tay$ establishes a nominal limit for which preferential alignment can occur for a given $\ell/d$. Finally, we emphasise that this mechanism is present only in the regime of Taylor vortex flow.

\section{Conclusion}\label{sec:conclusion}
In this study, we investigated finite-size, neutrally buoyant, prolate ellipsoids (aspect ratio $\Lambda=4$) in Taylor--Couette flow. The explored flow regimes are governed by pure inner-cylinder driving and comprise the regimes: Taylor vortex, wavy vortex, and turbulent Taylor vortex flow. The fluid phase is simulated using DNS, whereas particles are represented through an IBM approach. Two particles size ratios were considered; $\ell/d=0.1$ and $\ell/d=0.2$, for volume fractions 0.01\% and 0.07\%, respectively. Here, $\ell$ denotes the particle major axis and $d$ the  gap-width.

Upon releasing the particles at initially random location and orientation, we observe, after a transient, various distinctive particle distributions (figure \ref{fig:position_r_z}). These distributions are according to the flow regime and particle spatial distributions  categorised in modes $(i)$ to $(iv)$  (see \S\,\ref{sec:spatial_distributions}).
Mode ($i$) to $(iii)$ are observed in the Taylor vortex flow regime. Here, mode $(i)$ corresponds to steady large orbits away from the core. Remarkably, for higher $\Tay$ in the Taylor vortex flow regime, particles get trapped in the vortex core. This particle distribution is denoted as mode $(ii)$ and it is the focus of this work.  Interestingly, the $\Tay$ range corresponding to mode $(ii)$ is different for  $\ell/d=0.1$ and $\ell/d=0.2$. Moreover, the particle concentration for mode $(ii)$ was observed to be much higher for $\ell/d=0.2$ than for $\ell/d=0.1$ (figure \ref{fig:pdf_rad_pos}). Mode ($iii$) is a transition in which mode $(i)$ as well as mode $(ii)$ are observed. Mode $(iv)$ corresponds to particle distributions in the wavy vortex regime and turbulent Taylor vortex regime. Here, particles distribute throughout the domain due to the instabilities in the flow. 

Furthermore, we find distinctive particle orientations for each mode. Let $\theta_z$ denote the angle between the particle axis of revolution and the local cylinder tangent. We find for mode $(ii)$  a sharp peak  around $\theta_z=0$ in the  pdf  $P(\theta_z)$  (figure \ref{fig:widthVsTa}).  The ability of particles to align is found to depend on three factors: Firstly, the  gradient in the flow. We find the most pronounced alignment for particles with $\ell/d=0.2$ to occur at $\Tay=4.2\times10^5$. This was observed to closely match the axial vorticity minimum  at the vortex core ($\Tay=6\times10^5)$. In comparison, the axial gradient at $\Tay=1\times10^5$ (mode $(i)$) is observed to be two orders of magnitude higher. From the particle Lagrangian dynamics a stable alignment was not observed (\cf figure \ref{fig:rotkinenta56e5}$a$). Secondly, 
the ability of particles to cluster.  Figure \ref{fig:rotkinenta56e5}($b$) indicates that a stable particle alignment (absence of rotational energy) only occurs once a particle is in the near vicinity of the vortex core. 
Lastly, the onset of instabilities in the flow. We observed in the transition from steady to wavy vortex flow, that particles spread throughout the domain (\cf figure \ref{fig:position_r_z}$e,i,j$). The corresponding spatial distributions (mode $iv$) become significantly flatter than the ones of  mode ($ii$),  \cf figure \ref{fig:orientations}.  

Our results collectively indicate that shear, large-scale structures induced by the curvature of the domain, particle shape, and particle size play an important interlinked role on the dynamics of the particles themselves. TC flow is known for its rich variety of flow structures. By adding particles, we observed that the particle dynamics alter significantly when the driving parameter $\Tay$ is varied.  

 \section*{Acknowledgements}
We wish to express our gratitude to H.W.M. Hoeijmakers for helpful comments on  our manuscript.
This work is part of the research programme of the Foundation for Fundamental Research on Matter with project number 16DDS001, which is financially supported by the Netherlands Organisation for Scientific Research (NWO). We acknowledge that the results of this research have been achieved using the DECI resource Kay based in Ireland at the Irish Center for High-End Computing (ICHEC) with support from the PRACE aisbl. R.J.A.M.S. acknowledges the financial support from ERC (the European Research Council) Starting Grant No. 804283 UltimateRB. This work was partly carried out on the national e-infrastructure of SURFsara, a subsidiary of SURF cooperation, the collaborative ICT organization for Dutch education and research.
 
 \section*{Declaration of interests}
 The authors report no conflict of interest.
\appendix

\section{Numerical method particles} \label{app:numMethod}
\subsection{Newton--Euler equations}
The dynamics of rigid particles are governed by the Newton-Euler equations
\begin{equation}\label{eq:translation}
\begin{aligned}
\dfrac{1}{r_p}\dfrac{d}{dt}(r_p^2\dot{\varphi}_p)&=\dfrac{6}{\pi} \int (\boldsymbol{\tau}\cdot\boldsymbol{n})_{\varphi} \ dS + F_{\varphi}\\
\ddot{r}_p &=\dfrac{6}{\pi} \int (\boldsymbol{\tau}\cdot\boldsymbol{n})_{r} \ dS + r_p \dot{\varphi}^2_p + F_{r} \\
\ddot{z}_p &=\dfrac{6}{\pi}
\int (\boldsymbol{\tau}\cdot\boldsymbol{n})_{z} \ dS
+ F_{z}.
\end{aligned}
\end{equation}
Equations in \eqref{eq:translation} are non-dimensionalised with the length scale $D_{\textit{eq}}$ (volume equivalent diameter of a sphere) and velocity of the inner cylinder $\omega_ir_i$.
The terms $F_\varphi$, $F_r$ and $F_z$ denote the collision force due to short range particle-particle and particle-wall interactions (see \S\,\ref{app:collForcesAndTorques} for details).
 
The term $\int (\boldsymbol{\tau}\cdot\boldsymbol{n})_{i} \ dS$  is computed as
\begin{equation}\label{eq:torqwithk}
 \int (\boldsymbol{\tau}\cdot\boldsymbol{n})_{i} \ dS\approx -\sum_{l=1}^{N_l} c_l V_l f^{l}_{i} + \dfrac{d}{dt} \int_V u_i dV + \int_Vk_idV \quad i= \varphi,r,z
\end{equation}
with $-\sum_{l=1}^{N_l}c_lV_lf^{l}_{i}$ is the force integrated over the shell of the particle. The ratio between the Lagrangian volume, $V_l$, and Eulerian volume, $V_e$, associated with a single Lagrangian marker with index $l$ is denoted as $c_l=V_l/V_e$. 
The hydrodynamic load in  \eqref{eq:torqwithk} has in contrast to Cartesian coordinates \citep[\cf][]{Breugem2012} an additional term $k_i$ stemming from the centrifugal component of \eqref{eq:momentum}. The term $k_i$ is derived by integrating \eqref{eq:momentum} over the volume of the particle and observing that from \eqref{eq:momentum} the terms $u_ru_\varphi/r$ and $- u_r^2/r$ are non-vanishing. Here,
 $k_i\equiv(k_\varphi, k_r,k_z)^T$ with $k_\varphi = u_ru_\varphi/r$, $k_r =-u_\varphi^2/r$ and $k_z =0$. A similar argument for the torque results in
\begin{equation}
\int_{\partial V} r \times (\boldsymbol{\tau}\cdot\boldsymbol{n}) dS = \dfrac{d}{dt}\int_V r \times\boldsymbol{u} dV + \int r \times \boldsymbol{f} dV + \int r \times \boldsymbol{k} dV.
\end{equation}


The orientation of the particle, described in a Cartesian frame, follows
\begin{equation}\label{eq:rotation}
I_p\dfrac{d\boldsymbol{\omega}^b_p}{dt} + \boldsymbol{\omega}^b_p \times (I_p\cdot \boldsymbol{\omega}^b_p) = R\cdot \int r\times ( \boldsymbol{\tau}\cdot\boldsymbol{n}) dS + R\cdot \boldsymbol{T}.
\end{equation}
and is solved in a local body frame \citep{AllenM.P.andTildesley1991}, by employing a quaternion description of the orientation. $R$ represents the rotation matrix, that converts the torque to the local body frame and is obtained via
\begin{equation}\label{eq:rotMat}
R =
\begin{pmatrix}
	 q_0^2+q_1^2-q_2^2-q_3^2 & 2(q_1q_2+q_0q_3) & 2(q_1q_3-q_0q_2) \\
	2(q_1q_2-q_0q_3) & q_0^2-q_1^2+q_2^2-q_3^2 & 2(q_2q_3-q_0q_2) \\
	2(q_1q_3+q_0q_2) & 2(q_2q_3-q_0q_1) & q_0^2-q_1^2-q_2^2+q_3^2
\end{pmatrix},
\end{equation}
with $q_i$ the four quaternion components describing the orientation of the particle. Then, after \eqref{eq:rotation} is solved for, the quaternions are updated for each particle with
\begin{equation}
	\dfrac{d}{d t} \begin{pmatrix}
	q_0 \\
	q_1 \\
	q_2 \\
	q_3
	\end{pmatrix}
	 = \frac{1}{2}\begin{pmatrix}
	q_0 & -q_1 & -q_2 & -q_3 \\
	q_1 & q_0 & -q_3 & q_2 \\
	q_2 & q_3 & q_0 & -q_1 \\
	q_3 & -q_2 & q_1 & q_0 
	\end{pmatrix}
	\begin{pmatrix}
	0 \\ \omega_x^b\\\omega_y^b\\\omega_z^b
	\end{pmatrix}
\end{equation}

The IBM, here, uses a Moving Least Squares (MLS) approach to perform the interpolation and spreading of forces \citep{vanella2009,Spandan2017}. The MLS approach is beneficial because the transfer functions formulated using MLS conserves momentum on both uniform and stretched grids. At the same time, MLS also retains reasonable accuracy for conserving the exchange of torque between the Eulerian and Lagrangian mesh on slightly stretched grids \citep{vanella2009,detullio2016}, which we employ in our study.

\subsection{Short range collisions - forces and torques}\label{app:collForcesAndTorques}
 The collision force and torque, $\boldsymbol{F}=\{F_\varphi,\ F_r,\ F_z\}$ and $\boldsymbol{T}$, respectively, are obtained via a lubrication and soft sphere model \citep{costa2015,ardekani2016}. Following \cite{costa2015}.
 Roughness is taken into account for the lubrication model at a thickness of $0.0005\, D_{\textit{eq}}$ \citep{costa2015}. Once particles overlap, the contact model takes over. The contact time-scale is set to $8\Delta t$ of the Navier-Stokes solver. The collision force is computed via sub-stepping with an incremental timestep of $\Delta t/50$. For each sub-step, an iterative scheme is used that converges once the criterion $|x_i^k - x_i^{k-1}|<\Delta z/100$ is met, with $x_i^k$  the rotated major axis of the ellipsoid in the inertial frame. Note that obtaining the shortest distance between an ellipsoid and a cylinder is a non-trivial problem. In appendix \ref{app:collision} we extend the iterative scheme from \cite{Lin2002} to solve this problem. In appendix \ref{app:coll_validations} we validate the collision model with data from available literature.

\subsection{Collision of ellipsoids with the inner and outer cylinder}\label{app:collision}
An important aspect of particle-particle or particle-wall collisions is the shortest distance between the two geometries. In case of nonspherical ellipsoid-ellipsoid interaction, a successful approach has been demonstrated in \cite{ardekani2016} by employing the iterative scheme from \cite{Lin2002} to compute the shortest distance. Here, it is shown how by introducing a ghost particle the same algorithm can be used to obtain the shortest distance between an ellipsoid and a cylinder.

\begin{figure}
\centering
\includegraphics[width=0.98\textwidth]{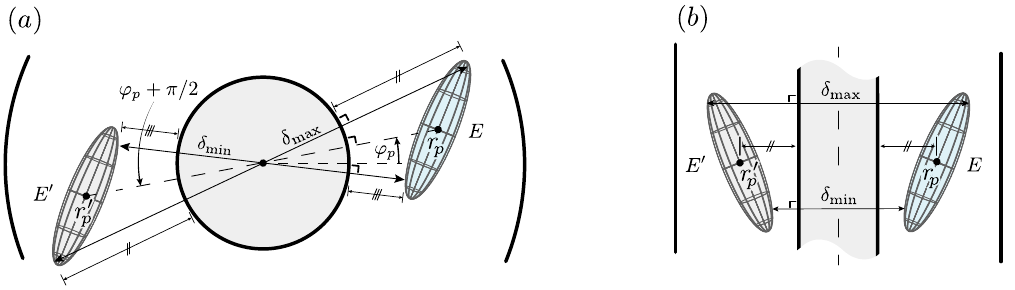}
\caption{Overview of a particle and its ghost particle for two different configurations (not at scale).   $\delta_{\min}$ and  $\delta_{\max}$ are used to calculate the shortest and longest distance between particle $E$ to the inner and outer cylinder respectively. $(a)$ Top view. $(b)$ Side view.}
\label{eq:particleShortestDistanceCylinder}
\end{figure}

 Given an ellipsoid $E$ for which we would like to calculate the shortest distance to the inner cylinder we introduce ghost particle $E^\prime$. The shortest distance is obtained by positioning $E^\prime$ such that the line of shortest distance connecting $E$ and $E^\prime$: (i) passes through the centreline of the inner cylinder and (ii) is perpendicular to the revolution axis of the inner cylinder. This is achieved as follows. Let $\varphi_p$ denote the azimuthal position of $E$. Then, $E^\prime$ is located at $\varphi_p+\pi$. Let $q =q_0 + q_1 i + q_2 j +q_3 k$ denote the quaternion of $E$ then  $E^\prime$ should be rotated such that $q^\prime = q_0 - q_1 i - q_2 j +q_3 k$. In figures \ref{eq:particleShortestDistanceCylinder}($a,b$) an overview is given for two different configurations, which illustrates the definitions for the distance computations.

The shortest distance $\delta_{\min}$ between $E$ and $E'$ can now be found in the conventional way as described in \cite{Lin2002}. The shortest distance from $E$ to the inner cylinder is then obtained via $\frac{1}{2}(\delta_{\min}-2r_i)$. 

The distance to the outer cylinder can be obtained by finding the largest distance $\delta_{\max}$ between $E$ and $E'$. One obtains $\delta_{\max}$ by picking the second root from the second order polynomial in the algorithm. Effectively, $\delta_{\max}$ is obtained by interchanging the $\min$ and $\max$ functions when computing the step-size \citep[\cf][(2.2)]{Lin2002}, \ie:

\begin{equation}\label{eq:stepsize}
\begin{aligned}
t_1 &= \min \{ t \in [0,1]: (1 -t)c_1 + t c_2 \in E\}\\
t_2 &= \max \{ t \in [0,1]: (1-t)c_1 + t c_2 \in E^\prime \}.
\end{aligned}
\end{equation}
 The particle distance to the outer cylinder is obtained via $\frac{1}{2}(\delta_{\max}-2r_o)$.

\subsection{Collision validations}\label{app:coll_validations}

Here, we validate our code similarly as performed in \citet{costa2015}. The validation consists of two parts. First, we compare the force exerted on a single particle approaching the inner cylinder and compare it to a theoretical prediction developed for a sphere approaching a flat plane at creeping flow conditions \citep{brenner1961,jeffrey1982}. The sphere has a diameter $D$ such that $D/x=16$ (uniform grid), and is positioned in a TC configuration with particle diameter to gap width ratio $d/D=12$. The inner cylinder to particle radius complies with $r_i/r_p=200$ and the particle Reynolds number is $\textit{Re}_p=0.1$. We position the sphere in a quiescent flow (cylinders are fixed) at a fixed distance and impose a velocity on its surface in the radial direction (Reynolds number $\textit{Re}_p=0.1$). We then let
the simulation run until a steady-state is reached and report the final value of the force $f_r$ acting on the sphere. 

\begin{figure}
 \centering
 \includegraphics[width=0.9\textwidth]{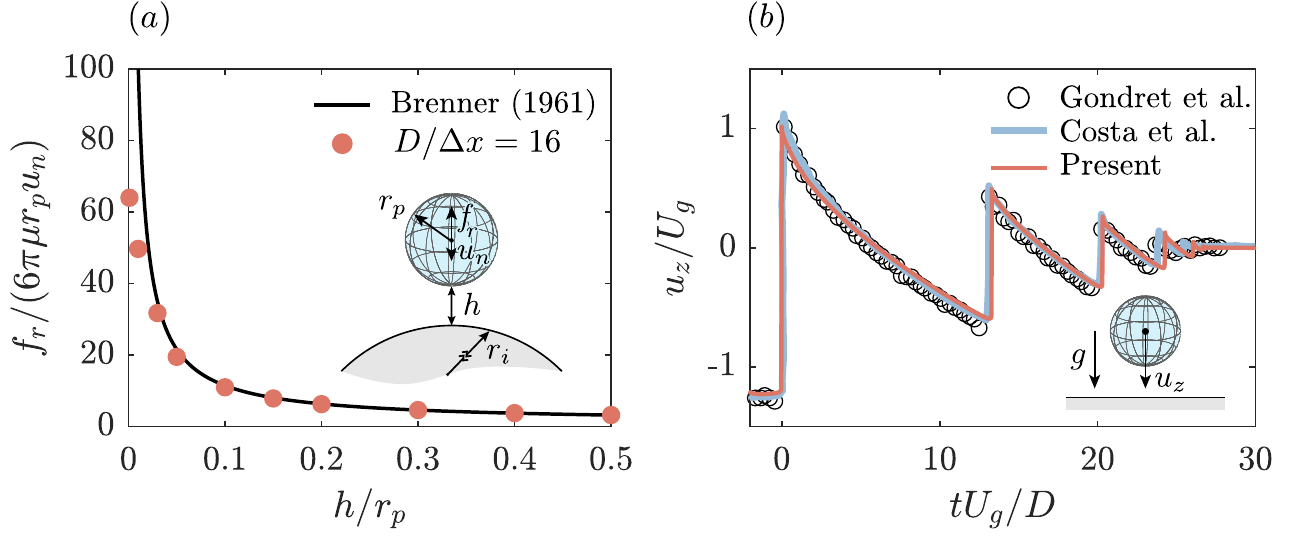}
 \caption{$(a)$ Normalised force exerted on a sphere approaching a cylinder versus the shortest distance between the geometries. The radius ratio is set to $r_i/r_p=200$.
 The theoretical predictions is from \citet{brenner1961}. $(b)$ A spherical particle falling on a flat plate at $\textit{Ga}=130.4$ and density ratio $\rho_p/\rho_f=8.34$. The results are compared with experimental data reported in \citet{gondret2002} and numerical results from \citet{costa2015}.}
 \label{fig:collision_validation}
\end{figure}

In figure \ref{fig:collision_validation}($a$) we report the shortest distance $h$ versus normalised $f_r$. The results  we observe are very similar to the results found in \citet{costa2015}, namely for distances of $h/r_p\gtrsim 0.03$ the numerical values of $f_r$ are in close agreement with the theoretical predictions. This figure shows that the simulation is not able to properly capture the asymptotic behaviour for $h/r_p<0.1$ due to the insufficient grid resolution between the sphere and cylinder.

The second validation is the replication of a sphere impacting on a flat wall, which bounces up and down until it is at rest. The non-dimensional parameters of the particle in relation to the fluid describing the problem are the Galileo number, $\textit{Ga}=130.4$ and density ratio $\rho_p/\rho_f=8.34$, with $\textit{Ga}=U_gD/\nu$. The characteristic gravitational velocity is defined as $U_g=\sqrt{|\rho_p/\rho_f-1|gD}$. The computational domain is $30D$ in vertical direction and $10D$ in horizontal direction. The top and bottom boundary are walls with zero velocity imposed on them. The side boundaries of the domain are periodic. 
A single sphere (uniform grid $D/\Delta x =16$ ) is then moved at the expected terminal velocity ($u_z/U_g=1.25$) for four particle diameters starting from the initial position of $29D$ above the bottom surface. From then on, the particle is allowed to freely fall. For a close comparison, we set the collision parameters equal to those reported in table II of \citet{costa2015}. Our findings of the particle velocity, $u_z$, are in close agreement to those reported in \cite{gondret2002} and \cite{costa2015} (see figure \ref{fig:collision_validation}$b$).

\section{Curvature effects}\label{app:curv_effects}
\begin{figure}
 \centering
 \includegraphics[width=0.7\textwidth]{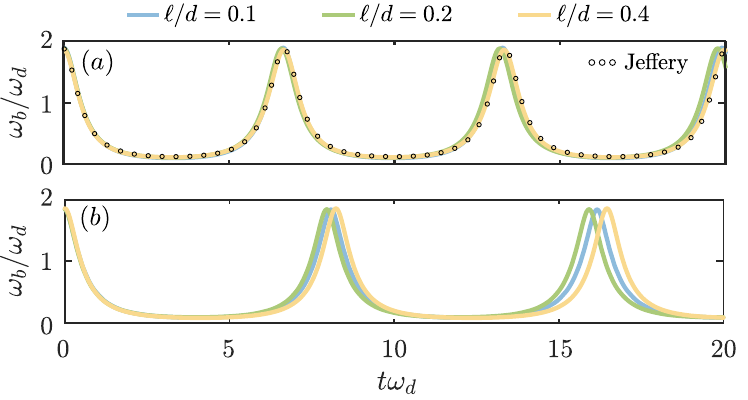}
 \caption{Curvature effects related to  particle size. The rotational velocity $\omega_b$ of a single pinned (where $u_\varphi =0$) particle is tracked. The quantity is normalised with $\omega_d=u_w/d$, with $u_w$ the wall velocity and $d$ the gap-width. $(a)$ Plane Couette flow. $(b)$ TC flow. For comparison the result derived by Jeffery is included \citep{Jeffery1922}.}
 \label{fig:curvature_effects}
\end{figure}

The effect of the TC curvature on the particle motion is examined here under conditions where Jeffery's equations are approximately valid. This is performed by comparing the motion of a single particle in plane Couette flow to the motion of a particle in TC flow. The TC configuration is taken as in the main study with a radius ratio of $\eta=5/7$. The inner and outer Reynolds number of the cylinders are $\Rey_i=-0.48$ and $\Rey_o=0.48$, respectively ($\Tay=1.0$). This flow is characterised as part of the circular Couette flow regime. Curvature effects are  assessed by varying the particle size. The particle is fixed at the location where the flow satisfies $u_\varphi = 0$. In comparison, the plane Couette geometry  is subject to the same  conditions with $Re_l=0.48$ and $Re_u=-0.48$ for the lower and upper wall, respectively. The particle is pinned at mid-gap.

The particle dynamics are presented in figure \ref{fig:curvature_effects}. The velocity gradient at the particle centre  is equal for both the TC system and plane Couette setup. Two observations can be made. First, the particle in TC is observed to rotate at a frequency $\omega_b$ that is $22\%$ slower compared to plane Couette. Second, the difference in rotation rate differs more between particle sizes for TC (difference  of $3.5\%$) in comparison to plane Couette ( $<1\%$). The higher difference  between different particle sizes for TC flow are understood as a curvature effect.


\end{document}